\documentclass[useAMS,usenatbib]{mn2e}
\usepackage{graphicx,epsfig,amssymb,amsmath,layout,verbatim,rotating,calc,mathrs
fs,natbib}
\usepackage{graphics}
\usepackage{rotate}
\usepackage{float} 
\usepackage{txfonts}

\input wasyfont

\newcommand{\dgr}{$^{\circ}$}
\newcommand{\ej}{E$_{J}$ }

\title[The phase space of peanuts]{The phase-space of boxy-peanut and
{\sf X}-shaped bulges in galaxies\\ II. The relation between face-on and edge-on
boxiness}
\author[P.A.~Patsis  \& M.~Katsanikas]
{P.A.~Patsis,$^{1,2}$\thanks{patsis@academyofathens.gr (PAP)}
M.~Katsanikas,$^1$\thanks{mkatsan@academyofathens.gr (MK)}\\
$^1$Research Center for Astronomy, Academy of Athens, Soranou Efessiou 4, GR-115
27, Athens, Greece\\
$^2$Excellence Cluster Universe, Boltzmannstr. 2, D-85748, Garching, Germany
}
\date{Accepted ..........Received .............;in original form ..........}

\begin{document}
\maketitle

\label{firstpage}
 
\begin{abstract}
We study the dynamical mechanisms that reinforce the formation of boxy
structures in the \textit{inner} regions, roughly in the middle, of bars
observed nearly \textit{face-on}. Outer boxiness, at the ends of the
bars, is
usually associated with orbits at the inner, radial 4:1 resonance region and can
be studied with 2D dynamics. However, in the middle of the bar dominate
3D orbits that give boxy/peanut bulges in the edge-on views of the models. In
the present paper we show that 3D quasi-periodic, as well as 3D
chaotic orbits sticky to the x1v1 and x1v1$^{\prime}$ tori, especially from the
Inner Lindblad Resonance (ILR) region, have boxy projections on the equatorial
plane of the bar. The majority of vertically perturbed 2D orbits, initially on
the equatorial plane in the ILR resonance region, enhance boxy features in  
face-on bars. Orbits that build a bar by supporting sharp ``{\sf X}'' features
in their side-on views at energies \textit{beyond} the ILR, may also have a
double boxy character. If populated, the extent of the inner boxiness
along the
major axis is about the same with that of the peanut supporting orbits in the
side-on views. At any rate these orbits do not obscure the observation of the
boxy orbits of the ILR region in the 
face-on views, as they contribute more to the surface density at the sides of
the bar than to their central parts. 
\end{abstract}


\section{Introduction}
\label{sec:intro}
Boxiness of galactic bulges is in most cases discussed in connection with the
boxiness observed in the edge-on views of disc galaxies. However, boxiness is
encountered also in nearly face-on views of barred galaxies. This can be either
an overall boxiness of the bar, usually referring to rectangularity of the
outermost isophotes of the bar component in early-type barred galaxies
\citep[see e.g.][]{ohw90}, or boxiness of the isophotes in the inner regions,
i.e. in distances from the centre roughly up to the middle of the semi-major
axis of the bar \citep{ed13}. 

As regards the outer boxiness it is plausible to assume that it is related with
orbits at the end of the bar, i.e. at regions with a much smaller scale height
than that of the regions of the bar that participate in the formation of the
peanut. In that respect, for the outer boxiness of the bar the presence of the
radial 4:1 resonance brings in the system rectangular-like p.o. \citep{gco88}
that attract around them quasi-periodic or sticky non-periodic orbits.
The role
of chaotic orbits in reinforcing the outer boxiness of the bars has been
indicated by \citet{w94}, \citet{kc96}, \citet{paq} and \citet{wp99}. This can
lead to the boxy shapes of early type bars. In the study of the outer face-on
boxiness of the bars, we deal practically with 2D dynamics. Thus, we find in 2D
models, that orbits associated with this feature are also related with the
orbits building the chaotic spirals beyond corotation \citep{p06, tp13}, i.e.
both structures share the same orbital content. Essentially these are orbits on
the equatorial plane. Nevertheless, the connection of the rectangularity of a
bar with the radial 4:1 resonance has been also detected in 3D models
\citep{psa03}. However, the relevant orbits remain at small heights away from
the equatorial plane.

For the \textit{inner} boxiness on the other hand we must use 3D orbits. If we
accept the
dimensions of a boxy bulge being as discussed in \citet{pI} (hereafter paper I),
the rectangularity of the inner isophotes \citep[as depicted in the
figures of][]{ed13} corresponds to regions occupied by the b/p bulge in the
side-on profiles. 
From this arises the question if there                               exist 3D
orbits contributing both to a face-on as well as to a side-on boxy profile. The
existence of this combination is not obvious. Thinking about the shapes of
periodic orbits, or about quasi-periodic orbits with similar shapes like the
periodic ones, this combination does not exist neither for this model nor for
any other of the models in \citet{spa02a, spa02b} or in the model used by
\citet{kp11}. The periodic orbits of the x1-tree (x1, x1v1, x1v2, x1v3 etc.)
follow in their projections on the equatorial plane the morphological evolution
of the x1 family as the Jacobi constant (hereafter called energy) increases. We
have to reach energies at the radial 4:1 resonance in order to obtain a
rectangular shape in their face-on, $(x,y)$ in our model, projection
\citep{gco83}. Orbits with such energies are found in the outer parts of the
bars. However, if we consider large $\Delta z$ or $\Delta p_z$ perturbations of
planar p.o. in the 4:1 resonance region, calculating in this way orbits with
large $|z|$ values, we enter into chaotic regimes of the models, where orbits do
\begin{figure*}
\begin{center}
\resizebox{160mm}{!}{\includegraphics[angle=0]{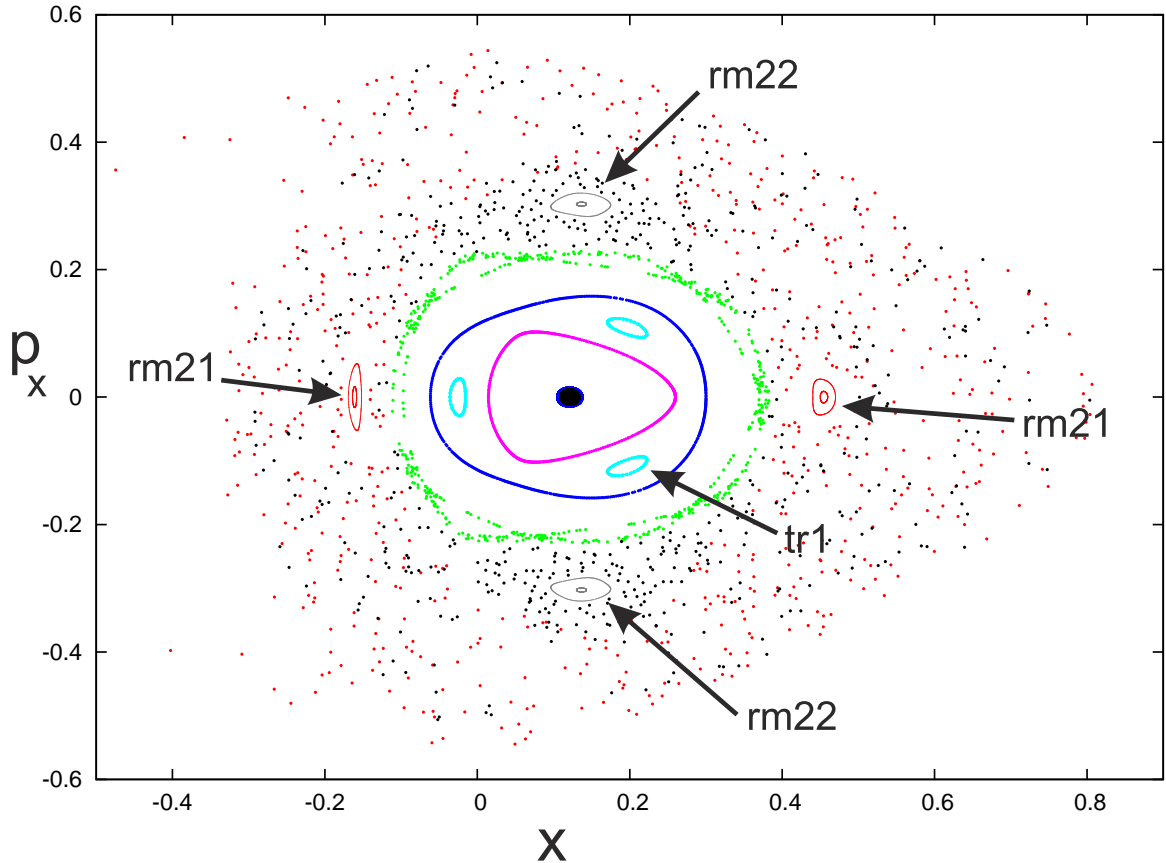}}
\end{center}
\caption{The $(x,p_x)$ cross section at \ej$=-0.41$, constructed by perturbing
the x1 initial conditions by $\Delta x$ (see text). The important for the
dynamics of our model families o multiplicity 2 rm21 and rm22, as well as the
family tr1, of multiplicity 3, are indicated with arrows. In this diagram  
the initial conditions of x1, x1v1, x1v2 and x1mul2 almost overlap at about
$(x,p_x)=(0.1,0)$ (heavy black dot).}
\label{xpx41}  
\end{figure*}  
not support any particular structure. By excluding orbits with large energies,
i.e. at or between the radial 4:1 resonance and corotation, we have to seek
after orbits in lower energies, where we encounter the orbits supporting the
boxiness of the edge-on profiles. 
Stable 2D orbits of the planar, ``3:1-type'', families could contribute to the
reinforcement of boxy structures on the equatorial plane \citep{psa03}.
Vertically perturbed orbits with $(x_0,p_{x_0})$ initial conditions on
their islands of stability are candidates for providing 3D orbits with boxy
projections on the $(x,y)$ plane. This possibility has to be examined for
representatives of these families with appropriate energies.
We already have seen that there are orbits
supporting the side-on b/p bulge without having a ``x1-like'' morphology in
their face-on projection (cf figure 19 in paper I). In the present paper we
examine the possibility of having among them orbits with boxy $(x,y)$
projections.

We use in our work the same model as in paper I and we keep the same formalism.
In section~\ref{sec:iboxi}, starting with 2D orbits on the equatorial plane, we
examine their orbital behaviour when they are perturbed away from it. Focusing
on the non-periodic orbits that support the b/p structure we first study their
dynamics in the typical case where all important families coexist and x1 is
stable. Then we investigate as well possible changes introduced in the energy
interval of the vertical ILR, where x1 is simple unstable. In
section~\ref{sec:3/1} we investigate the role of symmetric orbits bifurcated at
odd, radial n:1 resonances. Finally, we summaries our results and discuss our
conclusions in section~\ref{sec:concl}.


\section{3D orbits with boxy face-on projections}
\label{sec:iboxi}
\subsection{Dynamics at a typical energy}
An energy at which all main families of p.o. contributing to the b/p structure
coexist is \ej $= -0.41$.  At this energy x1, x1v1 and x1mul2 are stable, while
x1v2 is simple unstable \citep{spa02a, pI}. It is a typical energy for
understanding the dynamics at the peanut region. 
\subsubsection{x1 perturbations by $\Delta x$}
We examine the contribution of
non-periodic orbits in the neighbourhood of p.o. from these four families to the
reinforcement of inner boxy features, first by successively perturbing the x1
p.o. by increasing $p_x$. Fig.~\ref{xpx41} gives the $(x,p_x)$ cross section at
\ej $= -0.41$, which has been constructed this way. We  
investigate the phase space occupied by these orbits, when they
are kicked out of the equatorial plane. 
In the $(x,p_x)$ cross section the initial conditions of all
important families of p.o. involved in the enhancement of the boxy side-on
profiles, i.e. x1, x1v1, x1v2 and x1mul2, almost overlap at about
$(x,p_x)=(0.1,0)$. Their location is indicated with a heavy black dot in
Fig.~\ref{xpx41}. In the same figure we observe that the perturbed x1
orbits form around the p.o. a large central
region, characterized mainly by order. Inside this stability island we
discern three smaller stability islands belonging to a p.o. of multiplicity 3,
let us
call it tr1, and further out a chain of seven stability islands (drawn with
green) that mark the border between the ordered region and a chaotic zone that
surrounds it. Embedded in this chaotic zone we observe two sets of islands. They
are related to two p.o. of multiplicity 2, which we name rm21 and rm22 and are
centred at $(x,p_x)\approx(0.137,\pm0.3)$ and $(0.455/-0.16,0)$ respectively.
Arrows in Fig.~\ref{xpx41} indicate the locations of the orbits tr1, rm21 and
rm22. Other families of p.o. existing at this energy, like x2 and x3, are not
discussed at this point and are not indicated in Fig.~\ref{xpx41}. By
inspection of Fig.~\ref{xpx41} these three families of higher multiplicity seem
to play a minor role in the enhancement of any morphological feature in the
model. However, our investigation has shown the opposite, so we present here in
detail how they contribute to the enhancement of boxiness in both face-on and
side-on views. 

\begin{figure}
\begin{center}
\resizebox{80mm}{!}{\includegraphics[angle=0]{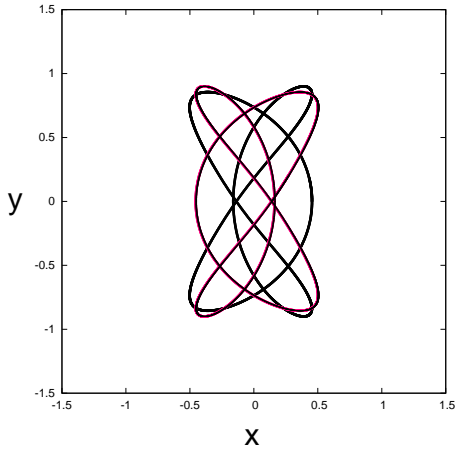}}
\end{center}
\caption{The morphology of the two planar simple unstable p.o. rm21 and rm22
(black and magenta) at \ej$=-0.41$.}
\label{rrorbs} 
\end{figure}

\begin{figure}
\begin{center}
\resizebox{84mm}{!}{\includegraphics[angle=0]{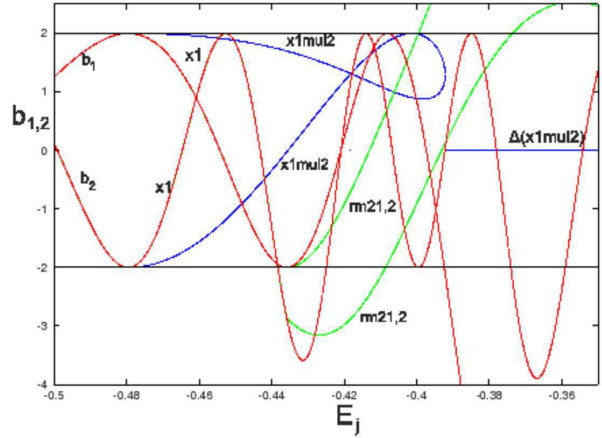}}
\end{center}
\caption{The red curves give the evolution of the stability indices of the x1
family when it is considered being of multiplicity 2. The family x1mul2 is
introduced in the system at \ej $\approx -0.479$.}
\label{mul2stab} 
\end{figure}

The families rm21 and rm22 have morphologies symmetric with respect to the $y=0$
axis. They are given together in Fig.~\ref{rrorbs}, with black and magenta
colours respectively. The combination of the two p.o. provide a morphology ideal
for supporting boxy structures in the central parts of face-on bars, first of
all on the equatorial plane. Even so, the system should provide for that purpose
also non-periodic orbits that will follow similar morphologies. In order to
trace the origin of the rm21 and rm22 families we plot the stability diagram of
x1 considering it as being of multiplicity two (Fig.~\ref{mul2stab}). Both rm21
and rm22 families have the same stability indices, as they are essentially two
branches of one family, symmetric with respect to the $x=0$ axis. They are
bifurcated from x1 at \ej $\approx -0.435$, where the b1 stability index of x1,
considered as being of multiplicity 2, has a tangency with the b$=-2$ axis. The
b1 stability index is associated with radial perturbations, so two 2D families
of p.o. are bifurcated. They are initially simple unstable (U), since the mother
family x1 is U at the energy they are introduced in the system. 

The other family of higher multiplicity we have encountered at \ej $= -0.41$ is
the orbit of multiplicity 3, tr1, which is indicated with an arrow, inside the
territory of the central stability island in Fig.~\ref{xpx41}. It coexists
together with its unstable counterpart of multiplicity 3, which is  located
between the three islands of stability (not indicated in Fig.~\ref{xpx41}). The
origin of tr1 can be found by considering it as a ``deuxi\`{e}me genre'' family
\citep{poin}. Since it is a radial bifurcation of x1 its starting point has to
be sought at an energy where the corresponding stability index of the simple
periodic mother family will be $b=-2 \cos(\frac{2\pi}{3})=1$ \citep[see e.g. the
appendix in][]{spa02b}. This happens for \ej $\approx -0.4617$ (cf figure
1b in paper I). 

\begin{figure}
\begin{center}
\resizebox{60mm}{!}{\includegraphics[angle=0]{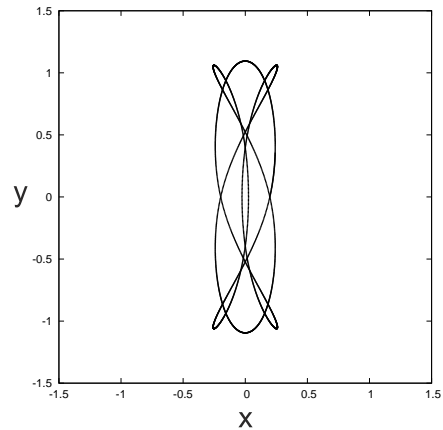}}
\end{center}
\caption{The stable family of multiplicity 3, tr1, at \ej$=-0.41$.}
\label{otr1} 
\end{figure}  
\subsubsection{Vertical perturbations of planar orbits on ``invariant curves''}
At \ej $\approx -0.4617$  tr1 is bifurcated from x1 as stable and extends
over a significant range of energies, including the ILR region, until it becomes
simple unstable at  \ej $\approx -0.4039$. Its face-on portrait is boxy, as can
be seen in Fig.~\ref{otr1}. In addition its stability secures motion on 4D tori
and sticky orbits on them. Perturbations up to $\Delta z=0.04$ of orbits
starting on the $(x,p_x)$ plane in the region between the dark blue and red
coloured invariant curves in Fig.~\ref{xpx41} give orbits on triple rotational
tori. The face-on morphologies of these orbits can be thought as thick versions
of the tr1 p.o. as depicted in Fig.~\ref{otr1}. Their side-on projections have a
b/p profile, however they remain close to the equatorial plane, reaching heights
$|z|\lessapprox 0.08$. 
\begin{figure*}
\begin{center}
\resizebox{160mm}{!}{\includegraphics[angle=0]{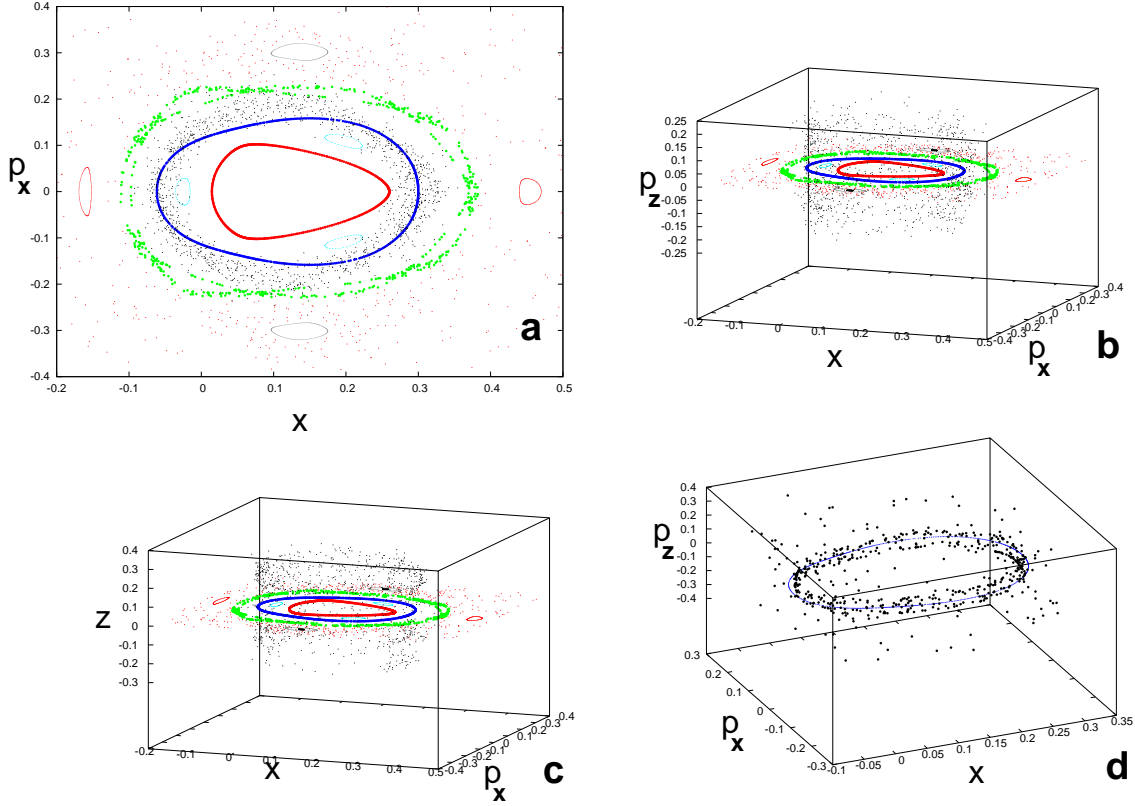}}
\end{center}
\caption{(a) The ring of black consequents projected in the region of the island
of stability in the $(x,p_x$) plane is formed by an orbit
starting on a closed curve around rm21. It stays trapped in this zone for 850
periods of x1 (\ej$=-0.41$). In (b) we give the 3D $(x,p_x,p_z)$ and in (c) 
$(x,p_x,z)$ projections of the space of section with the same orbits. We
observe that the consequents forming the ring in (a) stay confined in
cylindrical structures among the invariant curves of the central stability
island of the $(x,p_x$) plane in (a). In (d) we observe that also an orbit with
a small deviation in its initial conditions from the first one has a
qualitatively similar behaviour.}
\label{sticky41xpx} 
\end{figure*}  
Quasi-periodic orbits found by adding $p_{z_0} \neq 0$
to the orbits on the invariant curves in the $(x,p_x)$ plane give orbital
morphologies similar to those we find when we impose $\Delta z$ perturbations
with slightly thicker side-on views.

The closed elliptical curves around rm21 and rm22 indicate stability only on the
equatorial plane.  However, these families are at \ej$=-0.41$ unstable to
vertical
perturbations $(b_2=-2.15)$ and thus non-periodic orbits in their neighbourhood
in the $z$ or
$p_z$ directions will be chaotic. Even very small $\Delta z$ or $\Delta p_z$
perturbations of an orbit on an invariant curve around rm21 or rm22, bring the
perturbed orbits away from the $(x, p_x)$ plane. Nevertheless, we observe that
these orbits
during their integration stay confined in some regions of the 4D space of
section. As an example we consider an orbit starting \textit{on} an invariant
curve around rm21 with $(x_0,p_{x_0})=(0.47,0)$, which is perturbed by $p_z
=0.03$. We integrated it for time equal to 2000 x1 periods having an energy
conservation with
a relative error less than 10$^{-15}$. For time corresponding to 10 x1 periods
this orbit stays outside the region defined by the seven stability islands in
Fig.~\ref{xpx41}. Then its consequents are found inside
the seven islands region and they are projected in the $(x,p_x)$ plane around
the blue invariant curve. They stay there forming a ring for more than 850 x1
orbital periods. This is the ring of black points that can be observed in
Fig.~\ref{sticky41xpx}a close and around the blue invariant curve region. 

\begin{figure}
\begin{center}
\resizebox{80mm}{!}{\includegraphics[angle=0]{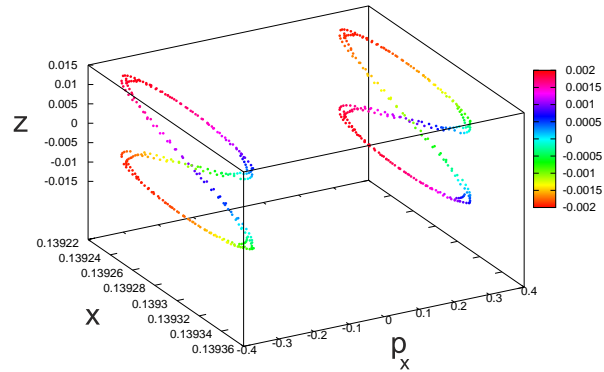}}
\end{center}
\caption{The 4D structure of a double tube torus around rm22. The
self-intersections of the ``8''-shaped structures is a projection effect, since
the two branches have different colours (colours are according to the $p_z$
values at the right hand side).}
\label{doubtor} 
\end{figure}  

The formation of a ring structure in the $(x,p_x)$ plane corresponds to a
spatial confinement of the orbits in both 3D projections of the 4D space of
section, which include this plane, namely $(x,p_x,p_z)$ and  $(x,p_x,z)$. In
both of these two 3D projections we have the formation of a cylindrical
structure that extends away from the $(x,p_x)$ plane remaining surrounded by the
chain of the seven stability islands. This can be seen in
Fig.~\ref{sticky41xpx}b and Fig.~\ref{sticky41xpx}c for the $(x,p_x,p_z)$ and
$(x,p_x,z)$ projections respectively. 3D projections including the $(z,p_z)$
plane are very illustrative for understanding the dynamical character of these
orbits and are described below extensively. For the time being we want to
underline that despite their sticky character these orbits are chaotic. In the
example of the orbit we describe, this means,
\begin{description}
\item[(i)] 
that if we extend the integration time, the consequents of the orbit will cross
again the border of the seven stability islands on the $(x,p_x)$ projection and
will
diffuse in a larger volume of the phase space. Nevertheless, later they  return
and stay trapped on the area of the ring of Fig.~\ref{sticky41xpx}a for hundreds
of x1 dynamical times. Also during the time an orbit spends away from a sticky
zone, i.e. in the example of our orbit away from the ring, it may stay confined
for a time interval within this period in another subspace of the space of
section. In our case the orbit stays for several hundreds of x1 periods
confined within
the area of the innermost, red, invariant curve in Fig.~\ref{sticky41xpx}a
before it returns again to the ring area and spends another time interval of
more than 200 x1 periods in this zone.
\item[(ii)]
that starting at nearby initial conditions we will have a quantitatively
different orbital behaviour. Nevertheless \textit{qualitatively} the overall
dynamical behaviour is similar. If we add a tiny $p_{x_0}=10^{-4}$ perturbation
to the initial conditions of the orbit with the black consequents in
Fig.~\ref{sticky41xpx}a,b,c we do not find the new orbit being trapped in the
same zones of the phase space during the same or close-by time intervals.
However, we find it spending time of the same order (hundreds of x1 periods) in
roughly the same zone, e.g. the ring area. The trapping of this new orbit in
the area around the blue invariant of Fig.~\ref{sticky41xpx}a is given in
Fig.~\ref{sticky41xpx}d.
\end{description} 

\subsubsection{Vertical perturbations of planar orbits in chaotic seas}
This orbital behaviour is typical not only for orbits starting on the invariant
curves of rm21 and rm22 in the $(x,p_x)$ plane, but for all orbits we started
integrating from the ``chaotic sea'' region of Fig.~\ref{xpx41}. This is to be
expected since essentially we deal with a single chaotic sea. In infinite time a
chaotic orbit will visit the whole available volume of the phase space. However
here, we integrate parts of this orbit starting from different initial
conditions. Even in the region $-0.409<$\ej\!$<-0.4$, where rm21, rm22 are
stable
(Fig.~\ref{mul2stab}) the above orbital dynamics does not practically change,
because the zone of influence around these stable p.o. is small. In this case we
have found a double ``tube torus'' \citep[i.e a self-intersecting torus in a 3D
projection, see][]{vetal97} in the 4D phase space. We present it in
Fig.~\ref{doubtor} at \ej$=-0.4084$ since it is the first time we find a tube
torus in a family of higher multiplicity. The tori belong to the perturbed by
$\Delta p_z =0.001$ rm22 orbit at that energy. The formed ``ribbons'' have an
intersection in the 3D $(x,p_x,z)$ projection, but not in the 4D space as one
concludes from the different colours (according to the $p_z$ coordinate) of the
intersecting branches. For slightly larger perturbations of the stable p.o. in
the $p_z$ or $z$ directions the orbits diffuse in phase space.
It is evident that the orbits originating in the chaotic sea around the central
stability region in the $(x,p_x)$ cross section (Fig.~\ref{sticky41xpx}a)
represent a major class of orbits, since they occupy a large volume of the phase
space and have similar behaviour in the examined range of energies
$-0.435<$\ej$<-0.375$. So, we want to investigate if there is a particular group
among them contributing to boxy structures in their face-on views. 
Their \ej values point to the right size of orbits for supporting the inner
boxiness we study. If so, we want also to find out what is the common feature of
such an orbital set. For example we want to investigate if they reinforce boxy
face-on structures when they stay at a particular region of the phase space. 

\begin{figure*}
\begin{center}
\resizebox{170mm}{!}{\includegraphics[angle=0]{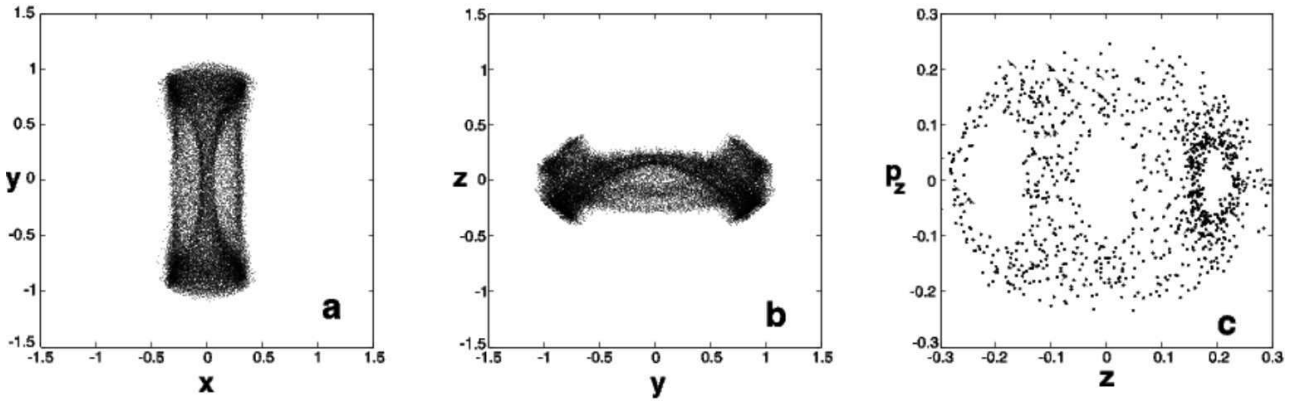}}
\end{center}
\caption{The morphology of the chaotic orbit during the time it remains trapped
in the ring region of Fig.~\ref{sticky41xpx}a, is given in (a) (face-on) and
(b) (side-on). The $(z,p_z)$ cross section of this orbit is given in (c). It
designates the stickiness of the orbit to the x1v1 and x1v1$^{\prime}$ tori.}
\label{upart} 
\end{figure*}

The first thing that we have examined is whether the orbit we have
studied starting on the invariant curve of rm21 on the $(x,p_x)$ plane, supports
a particular morphology during the time it spends in the ring inside the chain
of the seven stability islands. The result is striking as can be seen in
Fig.~\ref{upart}. 
In Fig.~\ref{upart}a we observe that the face-on projection has clearly a boxy
character. Since the orbit stays in the ring for more than 850 x1 periods we
sample it every 100 points during this time interval in order to
plot Fig.~\ref{upart}a. The dimensions of the box are comparable
with the dimensions of the tr1 p.o. in Fig.~\ref{otr1}. Simultaneously, we
observe that the side-on, $(y,z)$, projection of the orbit (Fig.~\ref{upart}b)
reflects the morphology of orbits trapped around x1v1 and x1v1$^{\prime}$, since
it appears as a combination of ``$\smile$'' and ``$\frown$'' parts. We already
know the imprint of sticky chaotic orbits to the tori of these families in the
$(z,p_z)$ plane from paper I (cf figure 14 in paper I). 
The spreading of the consequents of the orbit during the time of the 850 x1
periods in the $(z,p_z)$ reproduces exactly this typical pattern, as we can
observe in Fig.~\ref{upart}c.

We have already seen that during the time an orbit forms a ring of consequents
projected inside the area defined by the chain of the seven stability islands in
the $(x,p_x)$ plane, it can support an inner boxy feature in the face-on view of
the model. Next, as a second step, we examine if we have a similar behaviour in
the case of orbits having projected consequents during their integration time
only in the chaotic zone of Fig.~\ref{xpx41}. Such orbits exist indeed. As an
example we give the orbit with initial conditions $(x_0 ,z_0, p_{x_0},
p_{z_0})=(0.515,0.034,0,0)$, always at the typical energy \ej$=-0.41$. For time
corresponding to about 300 x1 periods, its face-on morphology has a boxy
character with dimensions similar to the combined rm21,2 p.o.
(Fig.~\ref{rrorbs}). This is given in Fig.~\ref{ob41}a. With black we depict the
orbit integrated for 10 periods of x1, while the red background depicts the same
orbit integrated for 300 x1 periods. The side-on morphology can be observed in
Fig.~\ref{ob41}b and it is boxy. Although not as sharp as the one in
Fig.~\ref{upart}b, this orbit has again a morphology similar to the one produced
by a combination of quasi-periodic orbits around x1v1 and x1v1$^{\prime}$. It
has an overall b/p morphology and it reaches heights close to $|z|=0.4$. Let us
call ``phase A'' the time interval corresponding to 300 x1 periods within which
the orbit has the morphology presented in Figs.~\ref{ob41}a and \ref{ob41}b. 
For time larger than 300 x1 periods, the morphology of the orbit changes in a
characteristic way and is given in Figs.~\ref{ob41}c and~\ref{ob41}d. The red
background corresponds now to the part of the orbit in the time interval between
300 and 1000 x1 periods after starting the integration. We call this period
``phase B''. With black we plot the orbit during the first 12 x1 periods during
``phase B''. The face-on view (Figs.~\ref{ob41}c) becomes rounder but evidently
it increases the surface density in two stripes parallel to the y-axis at
$x\approx \pm 0.35$. They can be better seen in the red background and are
indicated with arrows.  In the side-on view of the orbit (Fig.~\ref{ob41}d) we
realize that the denser stripes parallel to the y-axis are like two caps, that
restrict the extent of the orbit to $|z|\gtrapprox 0.4$. In intermediate heights
the orbit does not follow any particular morphology. 

The consequents of the orbit in Fig.~\ref{ob41}  remain projected
\textit{around} the central stability island of the $(x,p_x)$ plane
(Fig.~\ref{xpx41}). Practically they overlap with the chaotic sea as depicted in
Fig.~\ref{xpx41}. However, as we move from ``phase A'' to ``phase B'' they drift
from the area of the seven stability islands to larger distances. A few
consequents during ``phase B'' are found around the stability island of the
planar retrograde family x4. (The invariant curves around x4 are not drawn in
Fig.~\ref{xpx41}. They are located to the left  of the figure, reaching $x<0$
values outside the drawn frame). To a large extent the consequents of the orbit
in Fig.~\ref{ob41}  surround the stability region in Fig.~\ref{xpx41}
\textit{on} the $(x,p_x)$ plane forming two vaguely defined successive rings. 

\subsubsection{The dynamical mechanism}
\begin{figure*}
\begin{center}
\resizebox{117mm}{!}{\includegraphics[angle=0]{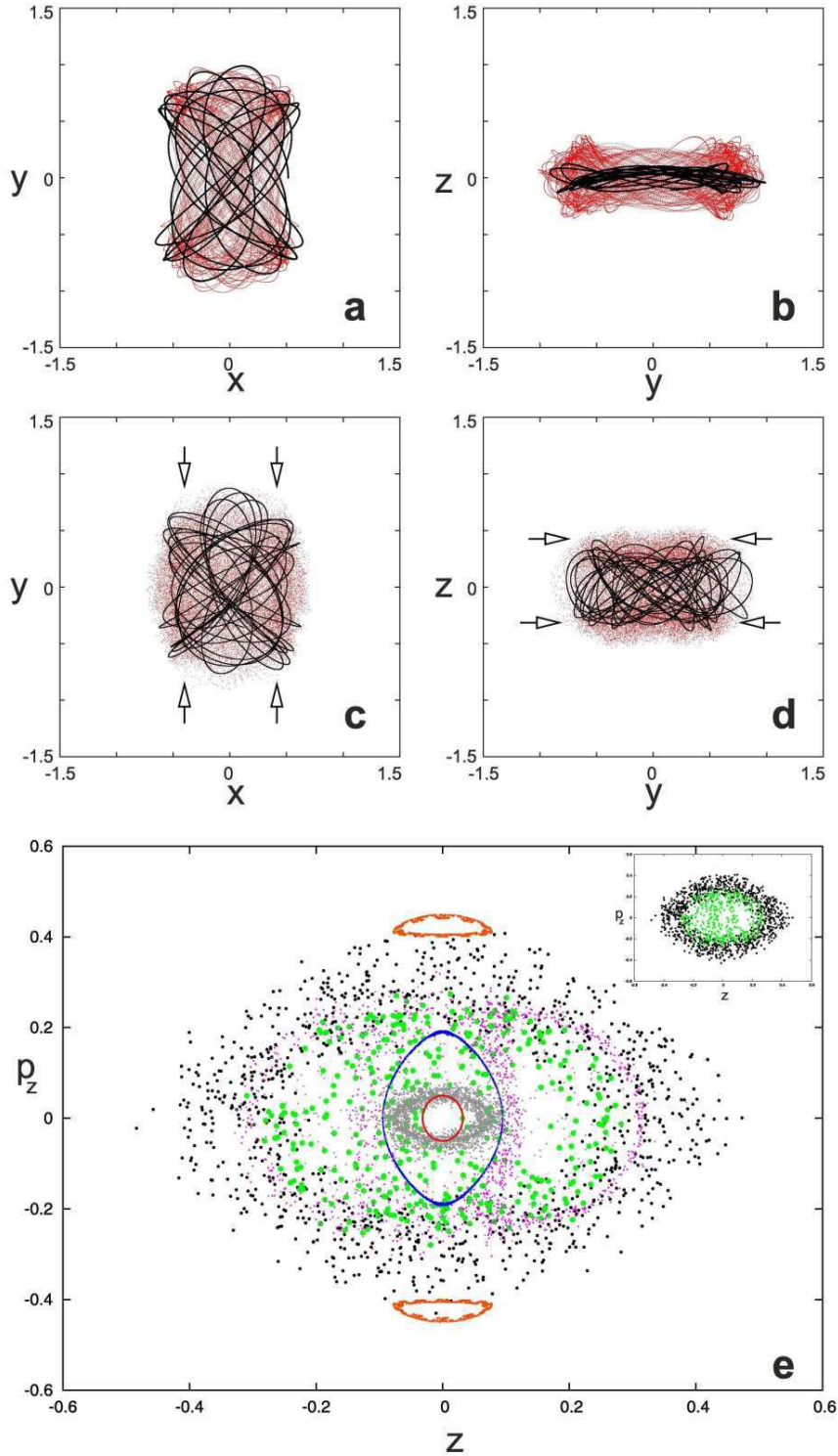}}
\end{center}
\caption{Contribution to boxy structures by an orbit with consequents projected
in the chaotic zone of Fig.~\ref{xpx41} (\ej $=-0.41$). In (a) and (b) we give
the face-on and side-on views respectively of an orbit starting at $(x_0 ,z_0,
p_{x_0}, p_{z_0})=(0.515,0.034,0,0)$ during phase ``A'' (see text), which lasts
for 300 x1 periods at the same \ej. With black we depict the orbit during the
first 10 x1 periods. In (b) and (c) we give the corresponding projections during
phase ``B'' (see text), in the course of the subsequent 700 x1 periods. In (e)
we present the $(z,p_z)$ plane of the consequents of this orbit. Green points
correspond to phase ``A``, while ''black`` to phase ''B'', while the inner grey
ring corresponds to a later phase, when the orbit becomes sticky to x4. The
magenta dots are the consequents of a sticky orbit to the x1v1 and
x1v1$^{\prime}$ tori. The central red and blue rings are projections of x1 tori.
In the embedded frame we isolate the consequents of the orbit during phase ``A''
and ``B'', so that it becomes evident that the consequents during phase ``B''
practically surround those during phase ``A''. Axes and dimensions of the
embedded frame are as for the main one in (e). }
\label{ob41} 
\end{figure*}

The projections of the consequents of the orbit of our example in the $(z,p_z)$
plane, are
given in Fig.~\ref{ob41}e. They are given together with the consequents of a
sticky orbit around x1v1 and x1v1$^{\prime}$  (magenta dots), as well as
together with the projections of the rotational tori of the x1mul2 family at the
top and the bottom of the figure around the points $(z,p_z)=(0.1,\pm 0.42)$ (cf.
figure 14 in paper I). The red and the dark blue rings in the central region of
the figure are the projections of one of the innermost and one of the outermost
tori around x1 respectively. Green points in  Fig.~\ref{ob41}e are the
consequents of the orbit during phase ``A''. They start being projected in the
$(z,p_z)$ plane inside the blue ring, but then they move outside it, and occupy
roughly the same region with the magenta orbit, which is sticky to x1v1 and
x1v1$^{\prime}$. The area covered by the consequents and their distribution on
the $(z,p_z)$ plane are very close also to the projected consequents of the
orbit in Fig.~\ref{upart}c. During ``phase A'' the orbit is unambiguously sticky
to x1v1 and x1v1$^{\prime}$. For longer integration times, during ``phase B'',
the consequents form a ring roughly surrounding the green consequents and are
plotted in black. Their relative location during the two different phases
``A'' and ``B'', are
given also in the embedded frame in the upper right corner of Fig.~\ref{ob41}e.
The dimensions of the embedded frame are the same as of the main frame of
Fig.~\ref{ob41}e. The ring of black points is located practically around the
green consequents and its extent in the $p_z$ direction seems to be hindered by
the x1mul2 tori. The weak confinement in two ring structures in the  $(x,p_x)$
and $(z,p_z)$ planes seems to be enough to secure a boxy character   in the
configuration space, of the kind described in Fig.~\ref{ob41}c and d.

The time intervals of hundreds of x1 dynamical times are already longer than the
time needed to consider a structure as supported by orbits. However, following
the orbit of Fig.~\ref{ob41} for even longer times gave further interesting
results. For time larger than 1000 periods of x1, the consequents in the
$(z,p_z)$ projection return inside the dark blue ring and form another ring,
this time coloured with grey in Fig.~\ref{ob41}e. The consequents remain trapped
on this ring for at least 2500 x1 periods. By the same time they remain trapped
on (rotational) tori around x4 and would be projected to the left of the
``chaotic sea'' in Fig.~\ref{xpx41} (roughly at $x < -0.2$). During this last
phase no boxiness is supported in the face-on view of the model. Instead the
orbit forms a counter-rotating disk on the $(x,y)$ plane. In the $(y,z)$
projection the orbit builds a thick layer close to the equatorial plane, without
any particular substructure. Thus, integrated for more than 1000 dynamical times
this orbit does not reinforce the double boxy profile. This case offers a
counter-example of a sticky orbit reinforcing a structure. Like quasi-periodic
orbits being trapped around a ``wrong'' stable periodic orbit, as is x4 for the
building of a galactic bar, during this phase the sticky chaotic orbit destroys
the boxy structure in our example. In the case of sticky orbits, our
example shows how the same  orbit can reinforce or destroy a
structure during different phases of its evolution. The information about the
trapping around different tori of p.o. can be obtained only directly by means of
the method of the surface of section. 

\begin{figure}
\begin{center}
\resizebox{82mm}{!}{\includegraphics[angle=0]{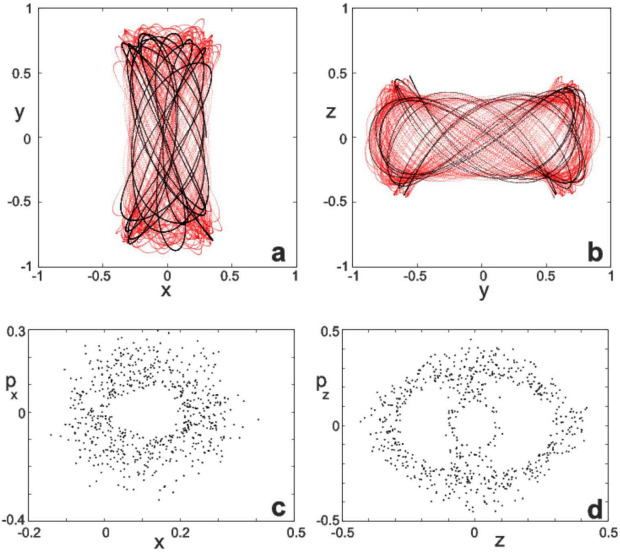}}
\end{center}
\caption{Typical morphologies of perturbed planar orbits with $(x_0,
p_{x_0})$ initial conditions in the central stability regions of
Fig.~\ref{xpx41}. In (a) and (b) we give the face-on and side-on views
respectively of an orbit starting at $(x_0 ,z_0, p_{x_0},
p_{z_0})=(0.3,0,0,0.3)$ and integrated for 100 x1 periods (\ej $=-0.41$). In
foreground with black we give the orbit during the first 10 x1 periods. The
$(x,p_x)$ and $(z,p_z)$ cross sections in (c) and (d) reproduce the
characteristic imprint of orbits supporting a double boxy character.}
\label{bb41stick} 
\end{figure}

Until now it became clear that a large number of planar orbits, which are
displaced from the equatorial plane by perturbations in the $z$ or $p_z$
directions become sticky to the x1v1 and x1v1$^{\prime}$ tori and during this
phase they support a double boxy morphology. This happens in the energy interval
$-0.435<$\ej$<-0.375$. However, the time interval during their integration,
during which they will be trapped at
specific regions of the phase space varies as we change the initial conditions.
Immediate support of the double boxiness from the beginning of the integration
is obtained by vertically perturbed orbits with $(x,p_{x_0})$ initial
conditions from the region between the dark blue invariant, just beyond
the tr1 islands and the orbit of multiplicity 7 in Fig.~\ref{xpx41}. By
perturbing initial conditions in this region in the $z$
or $p_z$ direction we find quasi-periodic orbits on 4D tori and orbits sticky to
these tori, which have a double-boxy morphology. As a typical example we give
the orbit with $(x_0,z_0,p_{x_0},p_{z_0})=(0.3,0,0,0.3)$. Its face-on and
side-on views are given in Fig.~\ref{bb41stick}a,b respectively. With black is
given the orbit integrated for 10 periods of x1 at \ej$=-0.41$, while with red
is the same orbit integrated for 100 x1 periods. However, the orbit continues to
reinforce the orbit for many hundreds of x1 dynamical times. 
We observe in Fig.~\ref{bb41stick}b, that the orbit reaches heights $|z|\approx
0.5$. The dynamical mechanism is similar to the one we described for the orbits
we started integrating in the chaotic sea of Fig.~\ref{xpx41}, namely stickiness
to the tori of x1v1 and x1v1$^{\prime}$. This becomes evident if we look at the
$(x,p_x)$ and $(z,p_z)$ projections of the consequents of the orbit of
Fig.~\ref{bb41stick}a,b.  They are given in Fig.~\ref{bb41stick}c,d. Especially
in the $(z,p_z)$ projection (Fig.~\ref{bb41stick}d) it is clear that the orbit
reproduces the pattern of the sticky orbits we have first encountered in figure
14 of paper I. The advantage of these orbits with respect to other orbits
reinforcing the double boxy profiles is that one obtains this morphology
immediately after starting integrating them without any delay.

\subsection{Dynamics in the vILR region}
\begin{figure}
\begin{center}
\resizebox{82mm}{!}{\includegraphics[angle=0]{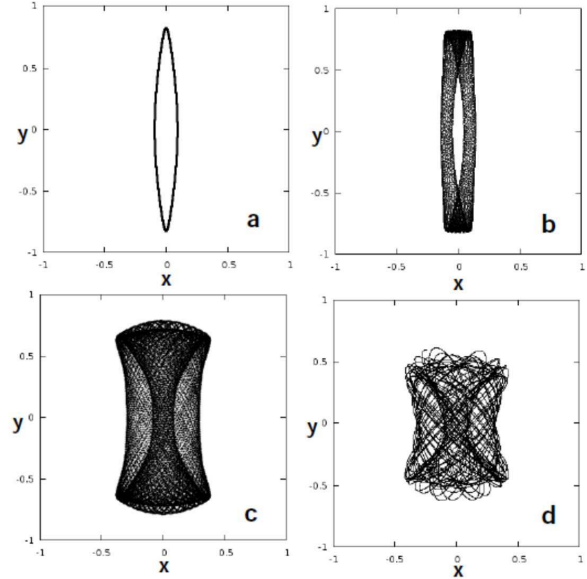}}
\end{center}
\caption{The evolution of the face-on projections of vertically perturbed orbits
in the energy interval in which x1 is simple unstable. All four orbits are for
\ej$=-0.438225$. (a) The perturbed by $z=0.02$ x1 orbit. (b) The perturbed by
the same $\Delta z$ orbit with $(x_0,p_{x_0},p_{z_0})=(0.14,0,0)$. In (c) we
increase the $x_0$ initial condition of the orbit in (b) to $x_0=0.29$ and in
(d) we increase the $z_0$ initial condition of the orbit in (c) to $z_0=0.25$.} 
\label{borbits} 
\end{figure}

In paper I we designated the energy interval in which x1 becomes simple unstable
at the vertical ILR as $\Delta$E$_{vILR}$. Unlike the typical case with
\ej$=-0.41$ we used to discuss the dynamical mechanism in the previous section,
in this interval x1 is not stable, but simple (vertically)
unstable. Nevertheless, in the $(x,p_x)$ plane we encounter around x1 invariant
curves.
We find that if we perturb in the vertical direction orbits on these
invariant curves we may encounter again boxy face-on
projections in the configuration space. The main reason for this is again the
proximity of the x1v1 and
x1v1$^{\prime}$ tori to the initial conditions of x1. There is always a $\Delta
z$ or $\Delta p_z$ range for which the perturbed planar orbits are affected by
the presence of these tori. The examples used  in paper I for discussing
vertically unstable x1 orbits, having 2D regular orbits around them in the
$(x,p_x)$ plane, are typical cases of such  orbits. The perturbed
by $\Delta z =
0.02$ x1 orbit at \ej$=-0.438225$, has a peanut shape side-on morphology (figure
5d of paper I). Its face-on view is a typical x1 ellipse with a negligible
thickness (Fig.~\ref{borbits}a). If we include $\Delta x$ perturbations in this
orbit and we start increasing them, the $(x,y)$ projections of the resulting
orbits in the beginning are thick ellipses. For $x_0 \gtrapprox 0.14$ (keeping
always $z_0=0.02$) their face-on projections have a clear boxy character
(Fig.~\ref{borbits}b). This boxiness increases with increasing $x_0$. For $x_0 >
0.19$ the face-on view of the orbits are as for the orbit in
Fig.~\ref{borbits}c, for which $x_0 \approx 0.29$. The face-on morphology
is boxy for all orbits starting with $x_0 \approx 0.29$, $p_x=0$ and
$0.01<z_0<0.26$. We give the orbit with $z_0=0.25$ in Fig.~\ref{borbits}d (the
orbits increase their projections on the x-axis as $z_0$ increases).  
It is worth noticing that despite being double-boxy (in face-on and side-on
views), and despite their similarity in their face-on projections, the  
orbits in Fig.~\ref{borbits}b, ~\ref{borbits}c and ~\ref{borbits}d, have
different boxy morphologies in their side-on views. The orbit in
Fig.~\ref{borbits}b has a shape like the one we have seen in figure 5d of paper
I. The side-on morphologies of the orbits in Fig.~\ref{borbits}c and 
Fig.~\ref{borbits}d are similar to other profiles we encountered in paper I (see
figures 6a and 6b respectively).

All these findings clearly show that there is a class of non-periodic orbits,
either quasi-periodic or (especially) sticky to quasi-periodic orbits, that
reinforce simultaneously boxy side-on and face-on profiles in our model. Such
orbits with energies from the radial and vertical ILR region of our model give
in their $(x,y)$ projections boxy orbits that reach distances about 1~kpc from
the centre along the y-axis. For larger energies, especially for \ej$ > -0.3$,
when x1v1 becomes again stable and other p.o. of the x1-tree are introduced in
the system, we find anew non-periodic orbits trapped by x1v1 tori. Such orbits
may have a double boxy character, but remain narrow and do not participate to
the building of the peanut. For example x1v3 or x1v5 orbits perturbed by $z$ or
$p_z$ can give orbits with boxy face-on profiles, which however remain in small
heights. These are orbits associated with a possible outer boxiness of the bar.

\section{Possible contribution from orbits bifurcated at odd resonances}
\label{sec:3/1}
As it is known \citep[e.g.][]{cg89} in rotating galactic potentials, at
the
radial n:1 resonances with n being an odd integer,
we have a type 3 bifurcation \citep[][section
2.4.3]{gcobook}. In that case, pairs of symmetric families are introduced in the
\begin{figure*}
\begin{center}
\resizebox{180mm}{!}{\includegraphics[angle=0]{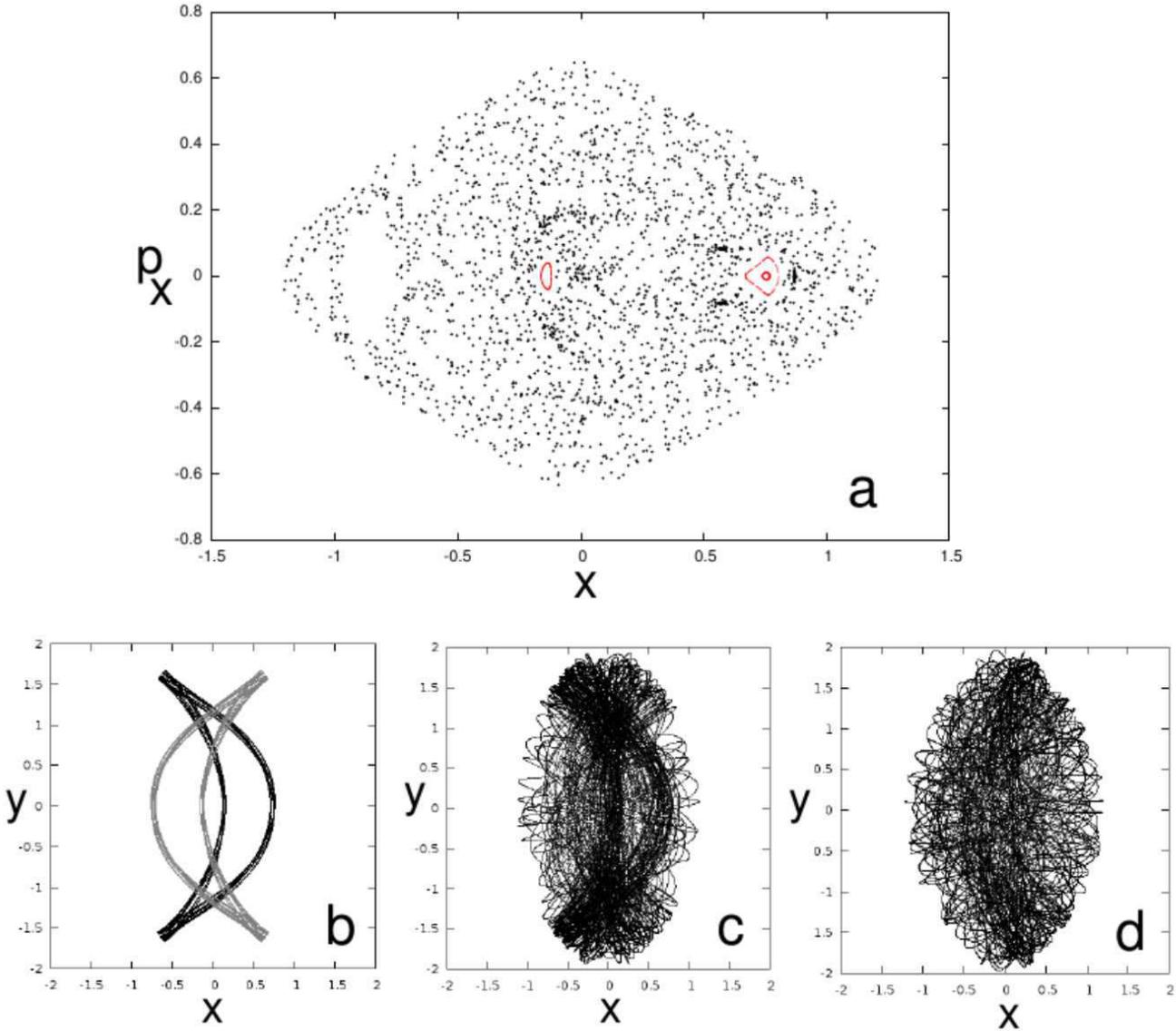}}
\end{center}
\caption{(a) A $(x,p_x)$ Poincar\'{e} section at
\ej$=-0.35$ constructed by perturbing the initial conditions of o1 by $\Delta
x$. The two small stability islands are around  o1 $(x>0)$ and o1$^{\prime}$
$(x<0)$. (b)
The $(x,y)$ projection of two quasi-periodic orbits around o1 and o1$^{\prime}$.
(c) A set of 13 sticky orbits with $z_0=0.2$ and $-0.2<x_0<1.0$ (d) A set of 10
sticky
orbits with $p_{z_0}=0.1$ and $-0.105<x_0<0.94$. The rest of the initial
conditions in all cases
are 0.  All
orbits are integrated for 10 o1 periods.}
\label{ot1fig} 
\end{figure*}  
system. Usually we encounter such families as bifurcations of x1 at  the 3:1
radial resonance. By plotting together both branches of the symmetric periodic
orbits belonging to the same family we create boxy features, as indicated by
\citet{psa03}. Recently, \citet{ch13} have presented chaotic orbits in the
neighbourhood of 3:1 p.o. that reinforce boxy morphologies close to the end of
the bar if integrated for long times. However, for the inner boxiness we study
here, the appropriate orbits have to support boxy features in the bar. In other
words their $(x,y)$ projections they have to lie inside the region occupied by
the corresponding projections of the 3D orbits that support the peanut in the
side-on view.  

In our model the 3:1 families are bifurcated close to the end of the
bar. Due
to their dimensions, possible reinforcement of boxy features by these orbits
will be associated boxiness at $y>2$. Also in the case of the potential from
an $N$-body simulation studied by
\citet{ch13}, such orbits contribute to the outer boxiness of the bar. However,
we have in our model another pair of planar symmetric
families, of 1:1 resonance type, which are found in the right energy range.
These orbits, called o1 and o1$^{\prime}$, are two branches of a
family bifurcated as stable from x1 at \ej$=-0.38$. They are discussed in
\citet{spa02b} and have a triangular type morphology. By considering both stable
symmetric branches of o1 \citep[see figure 17 in][]{spa02b} we construct a
skeleton that could support a boxy feature. In the interval $-0.38 < $\ej $<
-0.338$, o1, o1$^{\prime}$ are stable, x1 is (radially) simple unstable,
x1v1/x1v1$^{\prime}$  are complex unstable and x1v3 is simple unstable.
Fig.~\ref{ot1fig}a shows a $(x,p_x)$
Poincar\'{e} section for \ej$=-0.35$ constructed by perturbing the initial
conditions of o1 by $\Delta x$. We observe a main chaotic sea with two stability
islands belonging to o1 $(x>0)$ and o1$^{\prime}$ $(x<0)$. There is an increased
density of consequents around the islands, indicating the existence of sticky
zones. We followed again the procedure of perturbing initial conditions on the
$(x,p_x)$ plane by $\Delta z$ or $\Delta p_z$. We integrated them for time equal
to 10 o1 periods. We found that the quasi-periodic orbits trapped by either of
the symmetric branches of o1 have morphologies similar to those of the p.o.
when projected on the $(x,p_x)$ plane, but are 
thicker. The orbits in Fig.~\ref{ot1fig}b are the 
quasi-periodic orbits we obtain by perturbing o1, o1$^{\prime}$ by $\Delta z
=0.1$. 

Starting from the p.o., or from initial conditions on an invariant curve
around
o1/o1$^{\prime}$, or even from a $(x,p_x)$ point belonging to a sticky zone, we
find sticky orbits when exceeding a critical $\Delta z$ or $\Delta p_z$
perturbation. Typical examples are given in Figs.~\ref{ot1fig}c and
\ref{ot1fig}d. In Fig.~\ref{ot1fig}c we plot 13 orbits with
$(x_0,p_{x_0},z_0,p_{z_0})=(-0.2<x_0<1.0,0,0.2,0)$, while in Fig.~\ref{ot1fig}d
10 orbits with $(x_0,p_{x_0},z_0,p_{z_0})=(-0.105<x_0<0.94,0,0,0.1)$. Most of
these orbits are sticky to o1 and o1$^{\prime}$. We observe that the boundary of
these orbits has an oval shape. Combining sticky orbits we do not find symmetric
morphologies in general, even if we perturb by equal $\Delta z$ or $\Delta p_z$
the initial conditions of the periodic orbits. Non-periodic orbits associated
with the o1/o1$^{\prime}$ family fail to provide clear boxy features  in their
face-on views like the
orbits in Figs.~\ref{ob41}, ~\ref{bb41stick} and ~\ref{borbits}.  
By comparing 
the phase space in the case of the latter orbits with that of the orbits in
Fig.~\ref{ot1fig}, we observe that the main difference is the lack of the
rotational tori around x1v1 and x1v1$^{\prime}$. When o1 is introduced in the
system x1v1 and x1v1$^{\prime}$ are already complex unstable. This indicates
that the main dynamical phenomenon for having boxiness in the middle of the bar
is stickiness to these tori. In other models in which 3:1 type families exist at
the energies we have stable x1v1 and x1v1$^{\prime}$ orbits they may play a role
similar to that of rm2 and tr1 in Figs.~\ref{xpx41} and ~\ref{ob41}.

\section{Discussion and Conclusions}
\label{sec:concl}
Summarizing we can say that as long as a non-periodic orbit in the radial and
vertical ILR region of our model is trapped by, or becomes sticky to, x1v1
and/or x1v1$^{\prime}$ tori, it may visit a region of phase space, where the
orbits have rectangular-like face-on $(x,y)$ projections. Rectangularity in the
morphology of non-periodic orbits is evidently related with the presence of 2D
families of p.o. with boxy shapes in the equatorial plane. The presence of
families like rm21, rm22 and tr1 in $(x,p_x)$ cross sections indicate the
prevalence of boxy shapes in large regions of the phase space over a large
energy interval. Inner boxiness is generated by 
perturbing initial conditions of planar orbits
at energies where we find orbits with rectangular shapes on the equatorial plane
and x1v1/x1v1$^{\prime}$ tori in the phase space. As a result the orbits
participating in the inner boxiness have a double boxy character (face-on and
edge-on). Among them those that reach larger heights above the equatorial plane
tend to become more square.

Boxy periodic orbits of higher multiplicity facilitate the prevalence of
boxy
morphologies, because they influence larger regions of the phase space. In a
typical case we have small stability islands with larger sticky zones around
them. Sticky chaotic orbits visit phase space regions that are far from each
other. Planar families of symmetric p.o. introduced
at odd radial n:1 resonances do not play an important role for the inner
boxiness in
our model. However, they can be significant in models where they co-exist in the
same energy intervals with the tori of x1v1
and x1v1$^{\prime}$.

\begin{figure*}
\begin{center}
\resizebox{180mm}{!}{\includegraphics[angle=0]{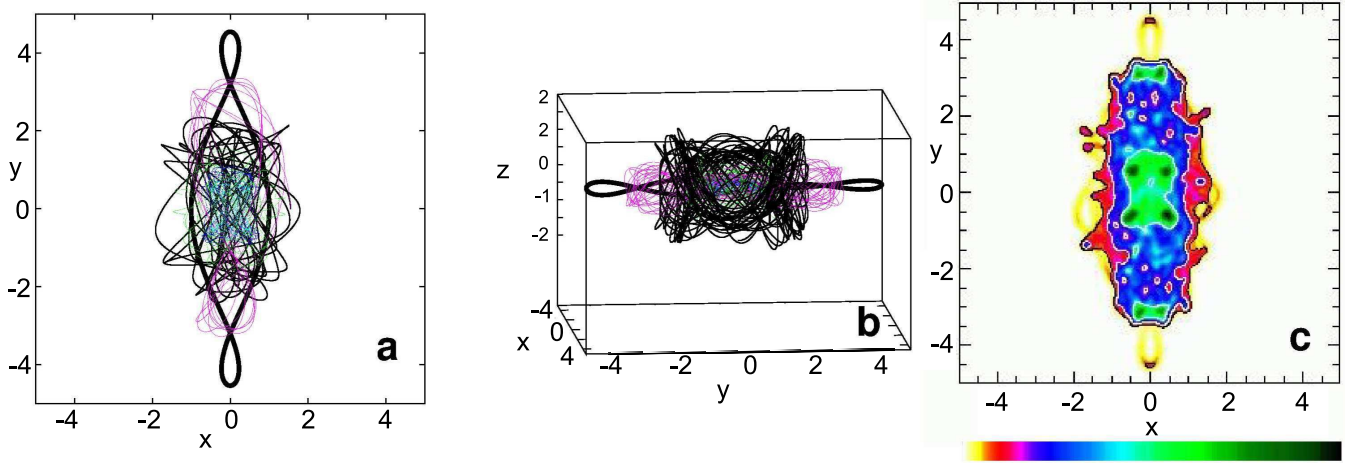}}
\end{center}
\caption{A set of representative non-periodic orbits that
support double boxy profiles. (a) The face-on and (b) a nearly side-on view of
these orbits. In (c) we apply a smoothing filter on an image based on the orbits
in (a) to indicate the boxy features reinforced in the model.}
\label{x1etot} 
\end{figure*}

\begin{figure}
\begin{center}
\resizebox{70mm}{!}{\includegraphics[angle=0]{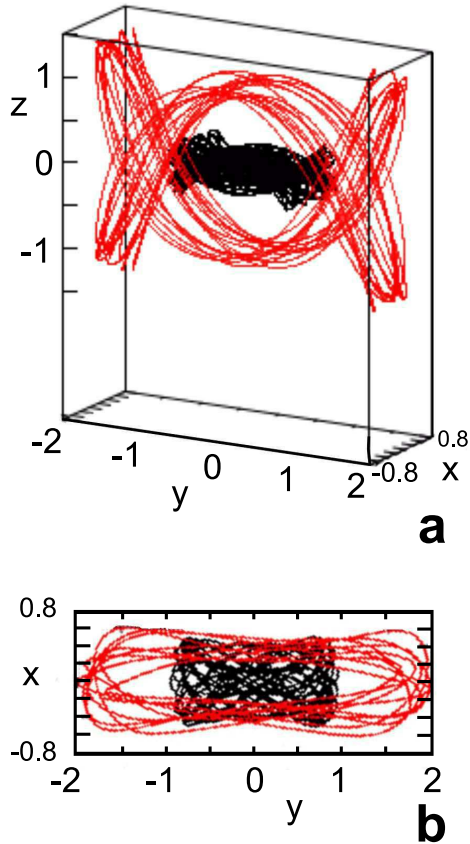}}
\end{center}
\caption{Two sets of orbits supporting the ``{\sf X}-feature'' having boxy
face-on projections both in a nearly side-on (a) as well as in their face-on
view (b).
The trapping of quasi-periodic orbits close to x1v1 and
x1v1$^\prime$ p.o. at the outer part of the b/p bulge (coloured red)
facilitates the observation of structures supported by orbits in lower heights
(coloured black) in the face on views of the model.}
\label{trans} 
\end{figure}

In Fig.~\ref{x1etot} we put together representative non-periodic orbits that
support double boxy profiles, both from the 2:1 resonance region as well as from
regions of higher resonances. The outermost, thicker drawn, black orbit,
discernible in Figs.~\ref{x1etot}a and b, is a x1v6 orbit that we identify with
the longest bar-supporting p.o. and is used as a measure of the length of the
bar in our model. The longest boxy orbit in Fig.~\ref{x1etot}a, drawn with
magenta colour is at \ej $=-0.254$. In Fig.~\ref{x1etot}b, which is an almost
side-on view of the same set of orbits, we can see the same magenta coloured
orbit remaining in much lower heights than the orbits that build the peanut. Its
detailed shape, hidden by the other orbits in the central regions of the model,
points to an association with x1v3. By converting Fig.~\ref{x1etot}a to an image
and by applying a smoothing filter, we obtain Fig.~\ref{x1etot}c. We can observe
how inner and outer boxiness can be formed in our model by means of
quasi-periodic and sticky orbits. It has to be underlined that orbits at higher
energies 
participating in the formation of the ``{\sf X}-feature'' in the side-on views,
especially by means
of orbits associated with the x1v1 and x1v1$^\prime$ families, contribute in
general less to the surface density of the central parts of the face-on views of
the bars and more to their periphery, as long as they retain a clear loop
character. The closer the ``{\sf X}''
supporting
non-periodic orbits remain trapped around the x1v1 and x1v1$^\prime$ periodic
orbits, the larger is the ``transparency'' towards the face-on rectangular-like
periodic orbits at lower energies with smaller heights above the equatorial
plane. This becomes evident in Fig.~\ref{trans} by means of a set of
quasi-periodic orbits with \ej$=-0.27$. Two are trapped around x1v1 and
x1v1$^\prime$ (coloured red), while the other two (coloured black) have boxy
face-on projections and are from the $\Delta$E$_{vILR}$ energy interval. In (a)
we have a nearly side-on and in (b) the face-on view. We observe that the outer
(red) orbits have also a rectangular-like face-on character, although not as
boxy as the inner two. 

In \citet{ed13} it is presented a list with 30 galaxies (see their A1,
A2
figures) with boxy isophotes in the middle of the bar. These are moderately
inclined galaxies, far from edge-on.
Since the bar of our model does not contain asymmetric terms, it is aligned with
the major axis of the system. So the extent and the proportions of the inner and
outer boxiness in the face-on views of the bar can be compared with rather
symmetric galactic bars, in galaxies with inclinations $i < $50\dgr like the one
of NGC~4037 or NGC~4123 \citep[see figures with isophotes in][]{ed13}.

\vspace{0.25cm} 
The main conclusions of our study are:
\vspace{-0.25cm} 
\begin{enumerate}
 \item The \textit{inner} boxy isophotes in the middle of the bars in nearly
face-on
views of barred galaxies can be supported by 3D quasi-periodic and mainly by 3D
chaotic orbits sticky to x1v1 and x1v1$^\prime$ p.o. in the ILR region. 
Inner boxiness is also characterized by the presence of planar periodic
orbits
with boxy morphology at the appropriate energies. By perturbing vertically such
p.o. we generate a class of
non-periodic orbits with double boxy character. The maximum distance along the
bar's major axis reached by these orbits is at or inside the limmits of the b/p
bulge.
Higher multiplicity p.o. influence large regions of the phase
space, mainly due to the presence of sticky zones around their stability
islands. Orbits sticky to multi-periodic orbits have consequents in the sticky
zones around different islands. This way a boxy structure can be supported
by orbits with initial conditions in regions of the phase space that are far
from each other.
\item At large
energies, orbits that reinforce ``{\sf X}-features'' in the side-on views of b/p
bulges by staying close to the periodic orbit, enhance locally boxy face-on
features without obscuring the sight towards boxy structures formed at lower
heights. The latter are located closer to the centre of the galaxies in the
$(x,y)$ projections.
\item Boxiness at the outer parts of the bars is supported by orbits associated
with families of higher resonances of the x1-tree, which remain very close to
the equatorial plane.
\end{enumerate}

\vspace{0.25cm} 
\noindent \textit{Acknowledgements}

We thank Prof. G.~Contopoulos for fruitful discussions and very useful comments.
This work has been partially supported by the Research Committee of the Academy
of Athens through the project 200/815.

\label{lastpage}

\end{document}